\documentclass[%
 aps,prl,
 amsmath,amssymb,
 reprint,%
superscriptaddress
]{revtex4-1}

\usepackage{graphicx}% Include figure files
\usepackage{dcolumn}% Align table columns on decimal point
\usepackage{bm}% bold math
\usepackage{siunitx}
\usepackage[caption=false]{subfig}

\newcommand{\Fig}[1]{Fig.~$\!${\ref{#1}}}

\newcommand{\comment}[1]{}

\usepackage{color}
\begin{document}

%\preprint{AIP/123-QED}

\title{Engineering Casimir interactions with epsilon-near-zero materials}% Force line breaks with \\
%\thanks{Footnote to title of article.}

\author{Miguel Camacho}
\affiliation{Department of Electrical and Systems Engineering, University of Pennsylvania, Philadelphia, PA 19104, USA}
\affiliation{Department of Electronics and Electromagnetism, Universidad de Sevilla, 41012 Seville, Spain}
 \author{Tao Gong}
\affiliation{Department of Electrical and Computer Engineering, University of California, Davis, CA 95616, USA}
\affiliation{Department of Materials Science and Engineering, University of California, Davis, CA 95616, USA}
\author{Benjamin Spreng}
\affiliation{Department of Electrical and Computer Engineering, University of California, Davis, CA 95616, USA}
 \author{Iñigo Liberal}
 \affiliation{Department of Electrical and Electronic Engineering, Public University of Navarra, 31006 Pamplona, Spain}
 \author{Nader Engheta}
  \email{engheta@ee.upenn.edu}
 \affiliation{Department of Electrical and Systems Engineering, University of Pennsylvania, Philadelphia, PA 19104, USA}
 \author{Jeremy N. Munday}
 \email{jnmunday@ucdavis.edu}
 \affiliation{Department of Electrical and Computer Engineering, University of California, Davis, CA 95616, USA}


\begin{abstract}
In this paper we theoretically demonstrate the tunability of the Casimir force both in sign and magnitude between parallel plates coated with dispersive materials. We show that this force, existing between uncharged plates, can be tuned by carefully choosing the value of the plasma frequency (i.e., the epsilon-near-zero frequency) of the coating in the neighborhood of the resonance frequency of the cavity. The coating layer enables a continuous variation of the force between four limiting values when a coating is placed on each plate. We explore the consequences of such variation when pairs of electric and magnetic conductors (i.e. low and high impedance surfaces) are used as substrates on either side, showing that this continuous variation results in changes in the sign of the force, leading to both stable and unstable conditions, which could find interesting potential applications in nanomechanics including nanoparticle tweezing.
\end{abstract}

\maketitle

The surprising existence of a force between two parallel metal plates in the absence of electric charges was predicted by Casimir in 1948 by means of an estimation of the rate of change of the zero-point energy associated with quantum electrodynamic fluctuations \cite{Casimir1948}. 

It was Lifshitz who several years later presented an exhaustive mathematical formulation for the calculation of such a force, which was already understood as being responsible for the attractive force between neutral molecular structures \cite{LIFSHITZ1992}. Due to the continuous frequency spectrum of the quantum fluctuations, the resulting force is given by a slowly-convergent integral over all frequencies, for which Lifshitz presents an expression in the form of an integral over imaginary frequencies with rapid convergence associated with the exponential decay of quantum electrodynamic fluctuations \cite{Ford1993,Parsegian2006, Ellingsen2008}. 

Due to the small amplitude of these forces, their experimental validation remained elusive for many years \cite{Derjaguin1978a}, and it was not until recently that their role in nanostructures with more complex interactions have been explored \cite{Klimchitskaya2009, rodriguez2011casimir,gong2021recent}. 

The ability to engineer its wideband frequency and wavevector net contributions launched the quest for materials that would allow for the manipulation of both magnitude and sign of the total net Casimir force \cite{Kenneth2002,Zhao2011}, including the use of complex artificial materials known as metamaterials \cite{Rosa2008}. In \cite{Munday2009}, it was shown that one can achieve repulsive Casimir forces for a wide distance range with naturally occurring materials. More recently, the combination of attractive and repulsive Casimir forces was achieved using multi-layered stacks to allow for stable trapping conditions in fluids, which could find applications in nanomechanics \cite{Zhang2019,ge2020gate}. It has also been demonstrated that anisotropic materials can lead to Casimir torques \cite{Somers2018}.

Most studies consider metallic materials acting as good conducting mirrors. Therefore, in such scenarios the plasma frequency, corresponding to the frequency point where the real part of permittivity crosses zero, occurs at frequencies that contribute little to the net Casimir interaction. However, the field of epsilon-near-zero (ENZ) optics, i.e., materials and photonic structures with a near-zero permittivity, has attracted a lot of attention due to its unusual wave effects \cite{Liberal2017a}. For example, these materials allow for special types of resonances that are able to selectively annihilate quantum fluctuations, which are in turn responsible for the Casimir force \cite{Liberal2017}.  

% \begin{figure}[htpb]
%  \centering
%  \includegraphics[trim=0cm 0cm 0cm 0cm, clip=true,width=\columnwidth]{Concept_3}
%  \caption{Geometry of the problem studied. A cavity between two perfectly conducting bodies is coated on either side with two distinct dispersive materials with thicknesses $d_1$ and $d_2$ respectively.}
%  \label{Fig_C}
%  \end{figure}

In this paper, we theoretically demonstrate the possibilities offered by the use of low-plasma-frequency materials for tailoring the Casimir force between high and low impedance surfaces (i.e. highly conducting or impeding surfaces). Specifically, we show that it is possible to achieve attractive and repulsive interactions with both stable and unstable equilibria that can be controlled by tailoring the plasma frequency, layer thickness, and/or surface separation.

\section{Coated electric conductors}
Let us consider two parallel perfect electric conductor (PEC) surfaces, each coated with a slab whose permittivities follow the Drude dispersion model given by

\begin{equation}
    \varepsilon_i(\omega)=1-\frac{\omega_{pi}^2}{\omega(\omega-i \omega_{ci})}
\end{equation}
with plasma frequencies $\omega_{p1}$ and $\omega_{p2}$ for the left and right coatings, respectively, and negligible losses ($\omega_{c1}=\omega_{c2}\approx0$) for an assummed harmonic time dependence $e^{i\omega t}$ . Although low loss has been assumed for simplicity here, small losses have been found to have little effect on the results presented here. 

As presented by Liftshitz, the Casimir pressure can be calculated in terms of an integral over the imaginary frequency $\xi=i\omega$ and real wavevector ${p}$, obtained through a rotation of the integration paths both in frequency and wavevector, given by \cite{Parsegian2006}

\begin{multline}
   P=-\frac{\hbar}{2 \pi^2 c^3}\int_0^\infty d\xi \int_1^\infty dp\ \xi^3 p^2 \\ \times\Bigg[\frac{{\Delta_1}\ {\Delta_2}e^{-r_n p}}{1-{\Delta_1}\ {\Delta_2}e^{-r_n p}} + \frac{\overline{\Delta_1}\ \overline{\Delta_2}e^{-r_n p}}{1-\overline{\Delta_1}\ \overline{\Delta_2}e^{-r_n p}}\Bigg] \label{forceq}
\end{multline}

\noindent where $\Delta_1$ and $\overline{\Delta_1}$ correspond to the reflection coefficients at the interface between vacuum and medium 1 (left) for transverse electric and transverse magnetic polarizations, respectively, and similarly for $\Delta_2$ and $\overline{\Delta_2}$ with respect to the right side of the cavity, and $r_n=2\ell\xi/c$ with $c$ representing the speed of light in vacuum, where $\ell$ is the vacuum-filled distance between the two surfaces. We follow the convention that a negative sign of the pressure corresponds to attraction, while a positive sign corresponds to repulsion.
 
\begin{figure}[htpb]
 \centering
 \subfloat[]{
 \includegraphics[trim=0cm 0cm 0cm 0cm, clip=true,width=\columnwidth]{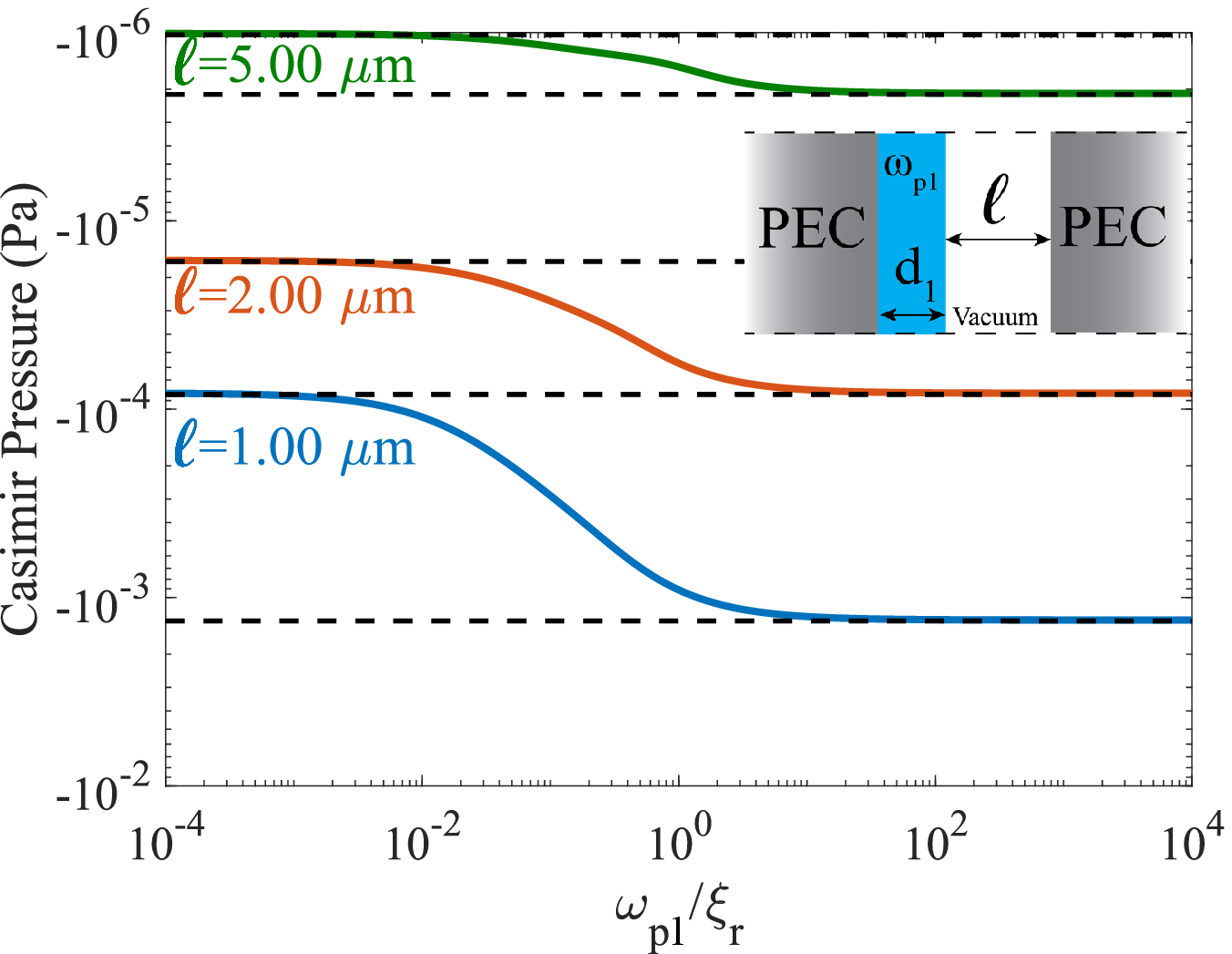}}
 \hfill
 \subfloat[]{
  \includegraphics[trim=0cm 0cm 0cm 0cm, clip=true,width=\columnwidth]{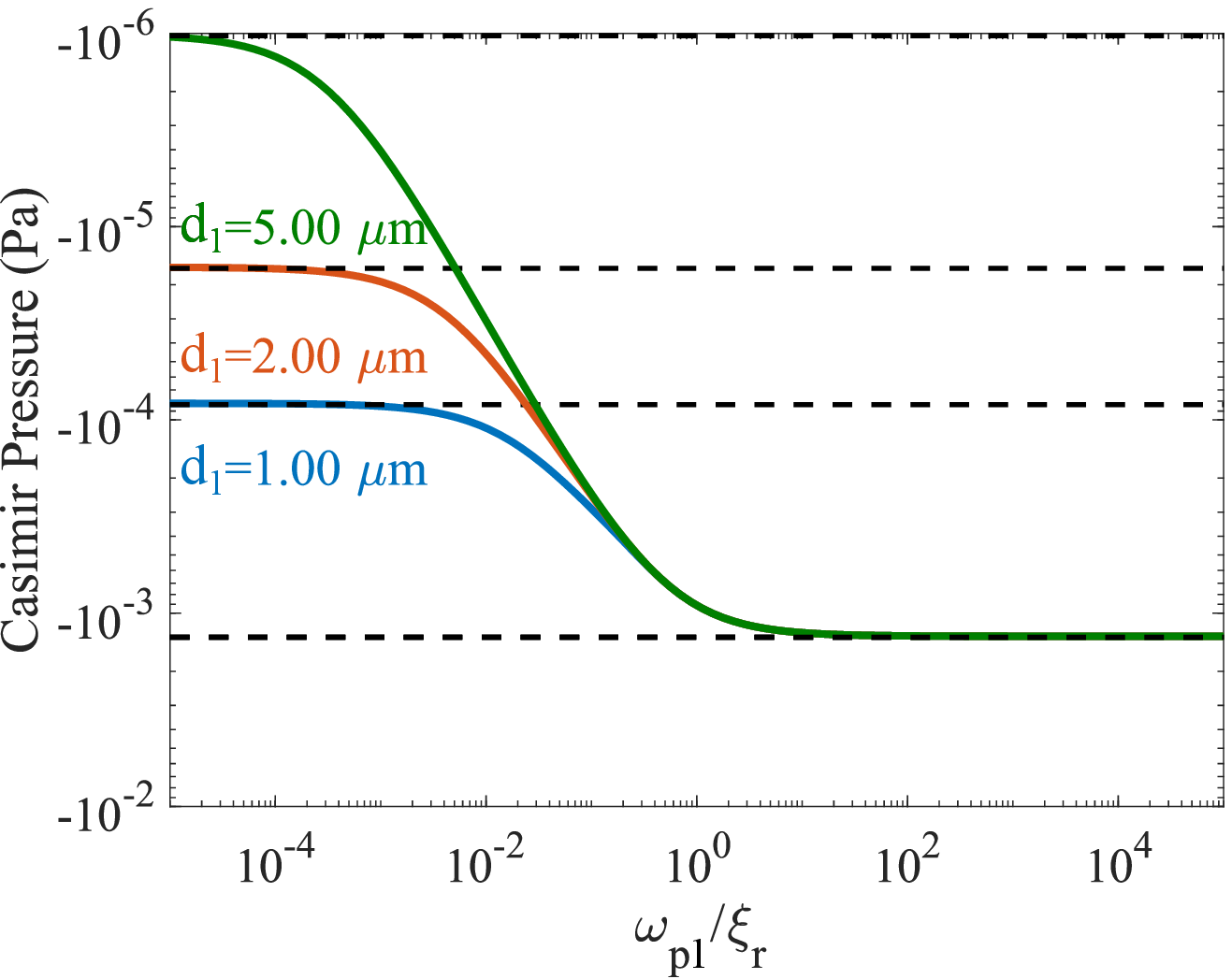}}
 \caption{Casimir pressure dependence on the plasma frequency (relative to the expected lowest resonance of the cavity formed by two highly conductive walls, $\xi_r$) for a cavity coated only on one side ($d_1=\SI{1}{\micro\meter}$ and $d_2=0$) for (a) three different distances, $\ell$, and (b) three different coating thicknesses, $d_1$ with $\ell=\SI{1}{\micro\meter}$.}
 \label{Fig_1}
 \end{figure}

As the starting point, we study the effect of changing the plasma frequency of one of the two coating layers, with constant thickness $d_1=\SI{1}{\micro\meter}$ in the absence of the second coating, i.e. $d_2=0$. We present these results in Fig.~\ref{Fig_1}(a), where one finds that as the plasma frequency is modified, the net attractive Casimir pressure (i.e. force per unit surface), is swept between two limiting values. These two limiting values correspond to the coating layer acting either as vacuum or as a perfectly conducting material for low and high values of the plasma frequency, respectively (given by the black dashed lines). This can be explained by the fact that the integral in \eqref{forceq} is dominated by the behavior of the materials involved around the imaginary frequency associated with the size of the cavity $\xi_r=c/\ell$. We show that this rationale holds for three different separations between the uncoated and coated plates. For the same reason, the maximal variation of the Casimir pressure takes place when choosing a plasma frequency near to that associated with the cavity size. In this manner, one can design the amplitude of the attractive Casimir interaction by tuning the plasma frequency of the coating layer.

In Fig.~\ref{Fig_1}(b) we complement the analysis by considering the case of constant separation between the interfaces on either side of the vacuum region (kept as $\ell=\SI{1}{\micro\meter}$) while we modify the thickness of the Drude-dispersive coating. We find that in this case, as expected by our rationale, the limiting value of the pressure associated with the highly-conductive regime of the coating is kept as a constant for the three cases studied. On the other hand, the second limiting value (associated with a vacuum-filled gap) is varied as the distance between the backing mirrors is increased consequently. In this case we find that, as the two limiting values become more distant, the transition region is increased and it provides the opportunity to tune in more detail the total pressure achieved using this multi-layer system. It can also be concluded from the figure that the pressure exhibits a monotonically increasing behavior with respect to the plasma frequency.

\begin{figure}[htpb]
 \centering
 \includegraphics[trim=0cm 0cm 0cm 0cm, clip=true,width=\columnwidth]{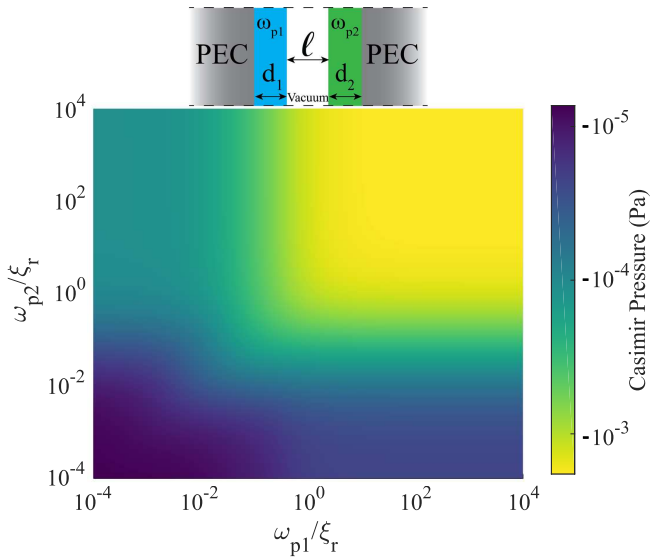}
 \caption{Pressure dependence on the two plasma frequencies (relative to the expected lowest resonance of the cavity formed by two highly conductive walls, $\xi_r$) for a cavity coated on both sides ($d_1=\ell=\SI{1}{\micro\meter}$ and $d_2=\SI{2}{\micro\meter}$) for a constant separation $\ell=\SI{1}{\micro\meter}$. }
 \label{Fig_3}
 \end{figure}

Let us now consider the effect of including a coating layer on both sides, for instance, by choosing $d_1=\ell=\SI{1}{\micro\meter}$ and $d_2=\SI{2}{\micro\meter}$. In this case, we can tune the plasma frequencies of the two coating layers, finding a more complex interaction. However, using the rationale presented before, we can expect to achieve four limiting values for the pressure, dictated by the combination of effective separations when considering none, either, or both of the coating materials as a highly-conductive (i.e. low-impeding) material. As shown by the results in Fig.~\ref{Fig_3}, by considering the possible combinations of plasma frequencies, we achieve all possible values in between these expected values of the pressure, which are in all cases attractive.

 \section{Coated low and high impedance surfaces}
 
With the understanding developed in the previous section for pairs of highly conductive plates coated with dispersive layers, let us now consider the use of high impedance surfaces, which act as perfect magnetic conductor (PMC) boundary conditions, imposing a zero tangential component of the magnetic field. These allow to modify the reflection coefficients involved in the calculation of the Casimir pressure in Eq.~\eqref{forceq}, which was shown in the literature to lead to attractive or repulsive Casimir pressures when different boundary conditions are combined \cite{Boyer}. For instance, both PEC-PEC (low-low impedance) and PMC-PMC (high-high impedance) cavities lead to attractive pressures, while a PEC-PMC (low-high impedance) cavity gives rise to a repulsive pressure. By adding an epsilon-near-zero coating into a PEC-PMC coating, we aim to manipulate both attractive and repulsive pressure components such that they can be balanced in a stable fashion. 
 
\begin{figure*}[htpb]
 \centering
  \subfloat[]{\includegraphics[trim=0cm 0cm 0cm 0cm, clip=true,width=\columnwidth]{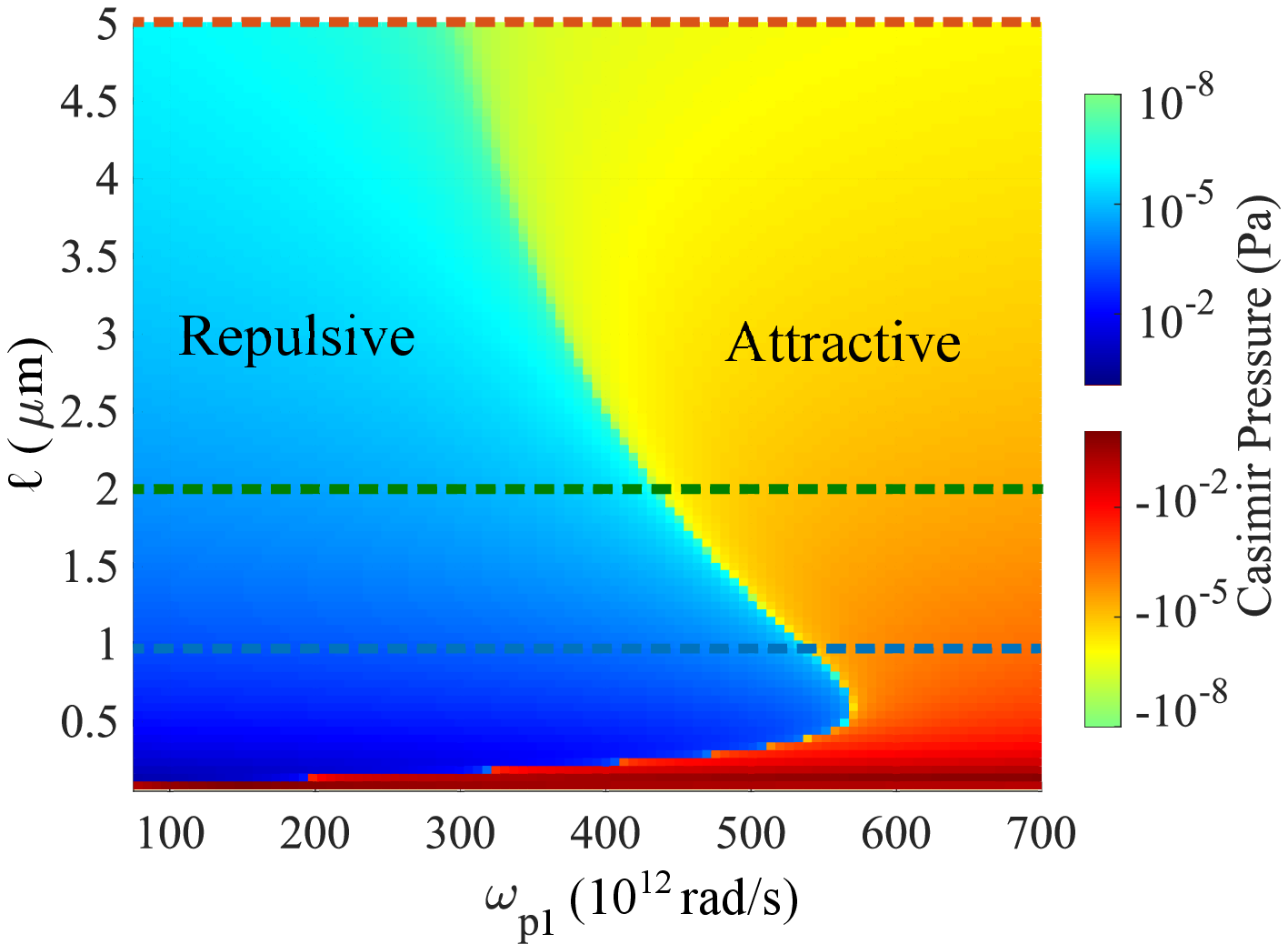}}
 \subfloat[]{
 \includegraphics[trim=0cm 0cm 0cm 0cm, clip=true,width=\columnwidth]{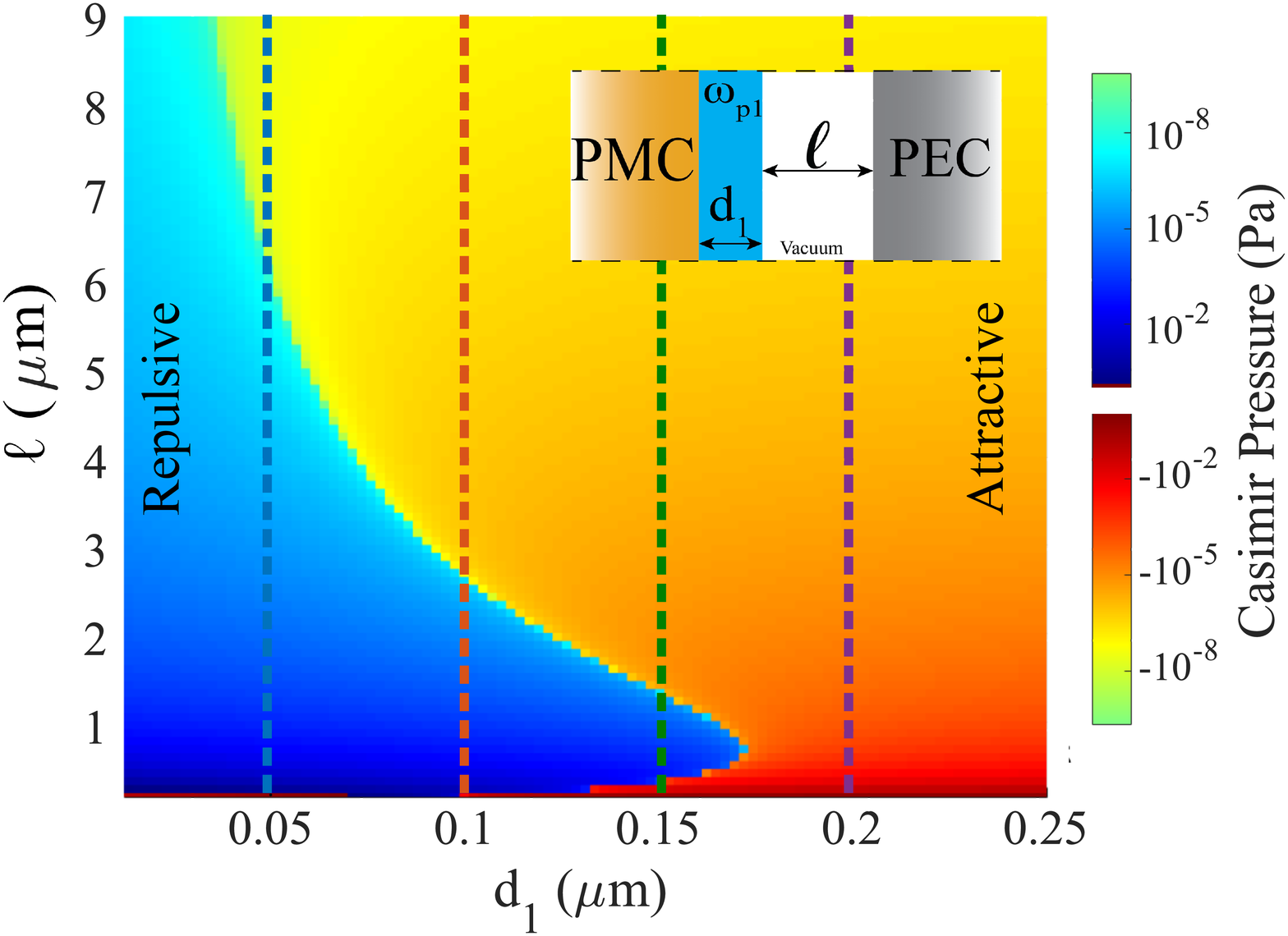}}
 \hfill
 \subfloat[]{\includegraphics[trim=0cm 0cm 0cm 0cm, clip=true,width=\columnwidth]{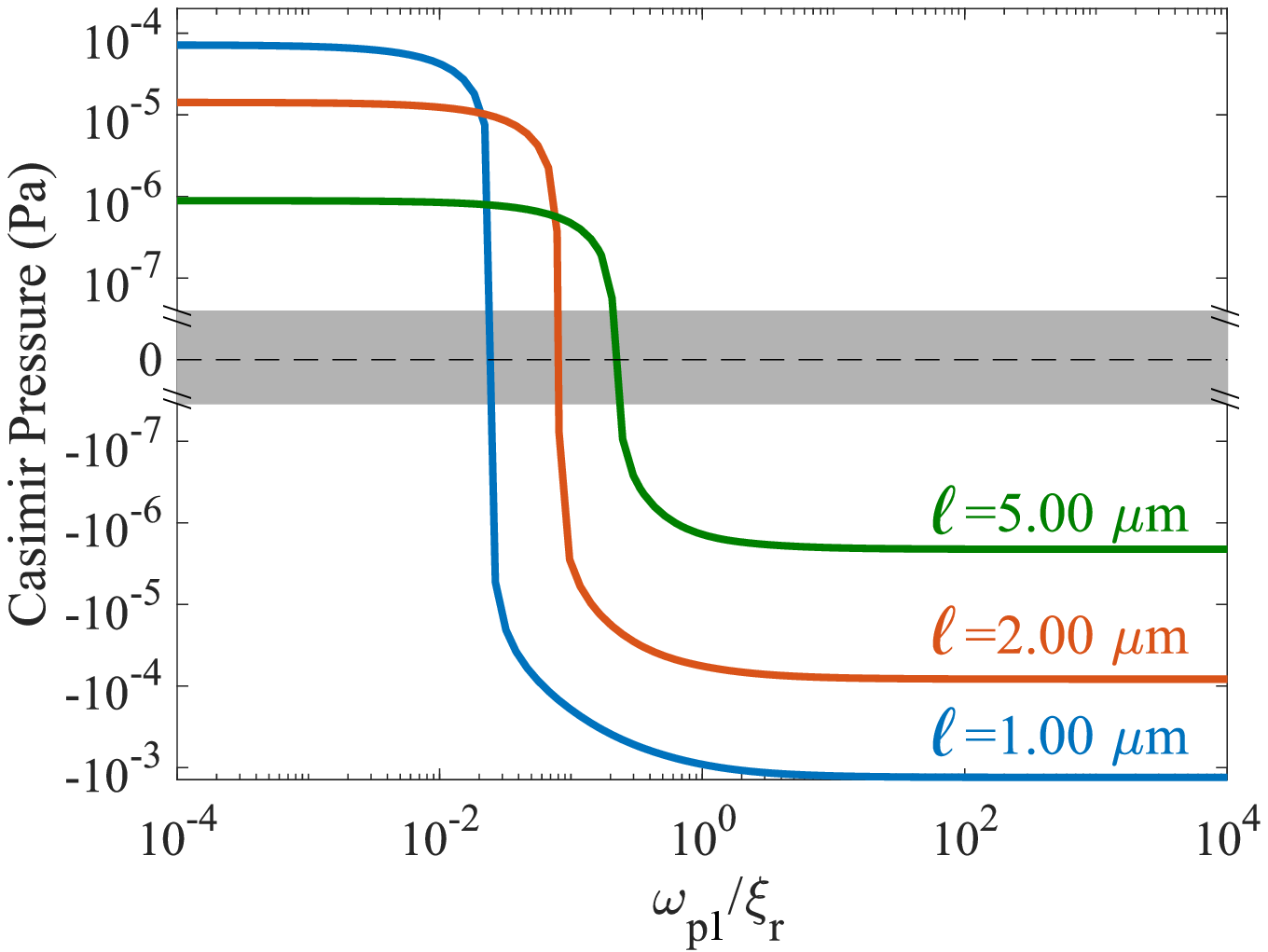}}
  \subfloat[]{
 \includegraphics[trim=0cm 0cm 0cm 0cm, clip=true,width=\columnwidth]{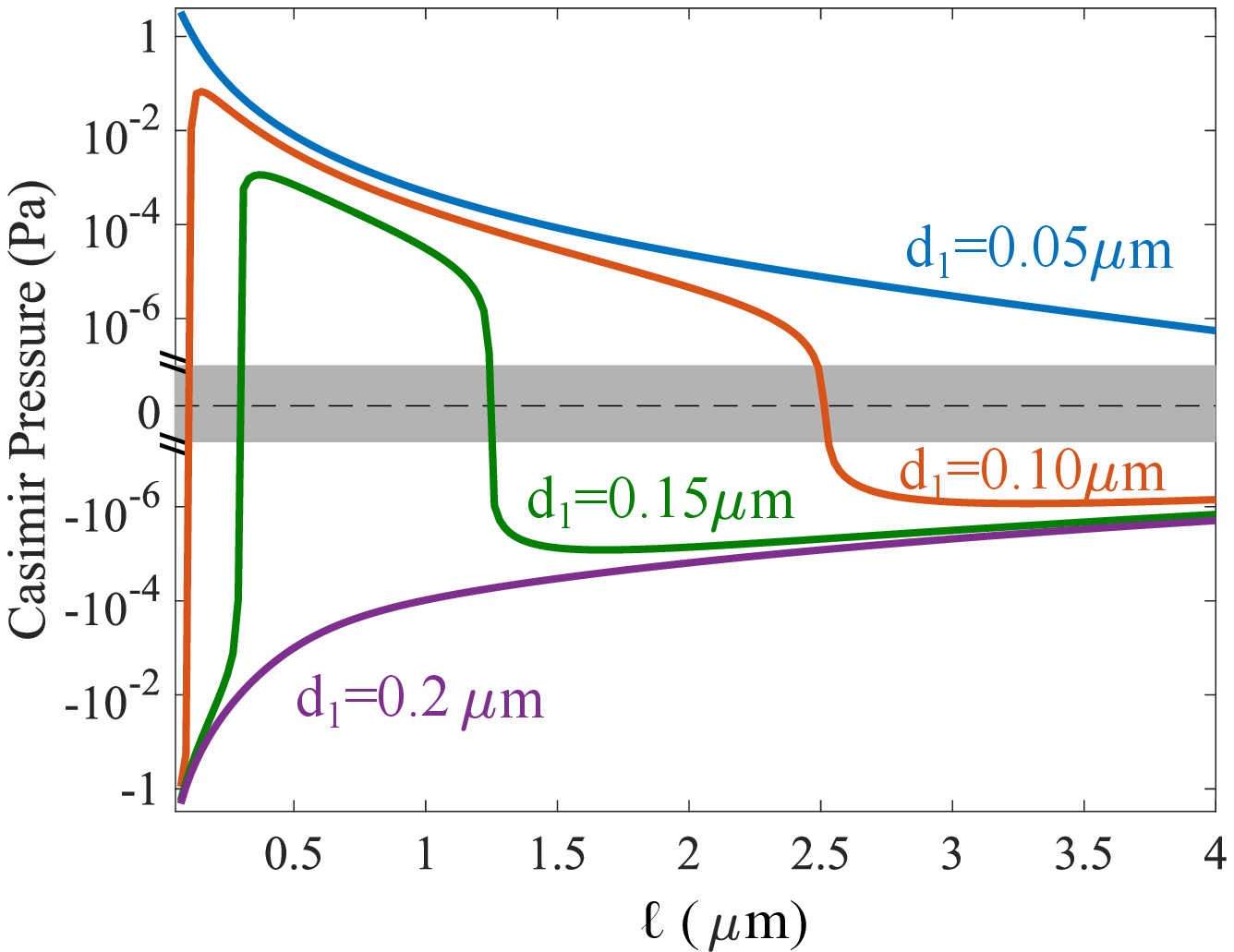}}
 \caption{Pressure dependence on the plasma frequency of the coating and separation for a cavity formed by a highly conductive wall and a coated non-dispersive high-impedance ($d_1=\SI{0.15}{\micro\meter}$ and $d_2=0$). Positive (negative) sign is associated with a repulsive (attractive) pressure. (a) shows the pressure dependence on both the cavity separation and plasma frequency of the coating for $d_1=\SI{0.15}{\micro\meter}$ and (b) shows the pressure dependence on both the cavity separation and coating thickness for a plasma frequency of $\SI{80}{\tera\hertz}$. (c) represents the extracted curves of the cuts shown in (a) and similarly between (d) and (b). }
 \label{Fig_4}
 \end{figure*}

Similarly to Fig.~\ref{Fig_1}, in Fig.~\ref{Fig_4} we consider the problem of a cavity formed by a high impedance wall on the left side and a zero-impedance wall on the right side, only the former being coated with a Drude-dispersive dielectric. For our analysis here, we assume the PMC to be nondispersive.  Using the physical picture presented earlier, one would expect that in the high-plasma frequency regime it would behave as a cavity formed by two highly conductive walls, therefore leading to an attractive Casimir pressure. However, in contrast to the previous section, now as the dielectric coating becomes more transparent, the waves encounter a high-impedance material, which leads to a repulsive pressure. 
 
Fig.~\ref{Fig_4}(a) shows the Casimir pressure map for a range of distances and values of the plasma frequencies. There we find the predicted crossing of the pressure through a zero value at a distance which is a function of the chosen plasma frequency. For a fixed distance, on either side of such crossing in terms of plasma frequency, we find the two pressures of different sign whose magnitude asymptotically approach those associated with either the smaller and larger cavities formed by the interfaces. Surprisingly, if one fixes the plasma frequency and varies the distance, it is apparent that there are two equilibrium points (one stable and one unstable) for a wide range of plasma frequencies up to $90\ \mathrm{THz}$, whose distance grows rapidly as the plasma frequency decreases.
 
As we show in Fig.~\ref{Fig_4}(c) by normalizing the plasma frequency to $\xi_r=2\pi c/\ell$ in each case, the zero pressure crossing appears at plasma frequencies close but below that associated with the size of the cavity. Additionally, such crossings occur within a small change of relative plasma frequency, and could lead to highly sensitive sensors for small displacements and material property variations. For example, with $\ell=\SI{1}{\micro\meter}$, half an order of magnitude change in $\omega_{p1}/\xi_r$ results in a 5 order of magnitude pressure change and with an order of magnitude change one achieves the switch from attraction to repulsion.
 
%  \begin{figure}[htpb]
%  \centering
%  \includegraphics[trim=0cm 0cm 0cm 0cm, clip=true,width=\columnwidth]{Fig4v2}
%  \caption{Force dependence on the plasma frequency (relative to the expected lowest resonance of the cavity, $\xi_r$) for a cavity formed by a highly conductive wall and a coated high-impedance ($d_1=\SI{1}{\micro\meter}$ and $d_2=0$) for three different distances $\ell$. Positive (negative) sign is associated with a repulsive (attractive) force.}
%  \label{Fig_4}
%  \end{figure}
 
To shed light on the parametric dependencies of the stability and instability points associated with the zero Casimir pressure, in Fig.~\ref{Fig_4}(b) we present the pressure dependence on the thicknesses of the vacuum and dispersive regions within the cavity formed by a perfect electric and a coated non-dispersive magnetic conductors. For that we have chosen a plasma frequency of $\SI{80}{\tera\hertz}$. In there, we find that the distance at which the pressure nulls does not vary monotonically with the thickness of the dispersive layer, which is a remarkable phenomenon. In fact, this effect allows us to find equilibrium points with both stability and instability conditions at different distances. For instance, we find a coating layer with a thickness $d_1$ of $\SI{0.15}{\micro\meter}$ would present a stable equilibrium point at $\ell\approx\SI{0.3}{\micro\meter}$ and an unstable equilibrium point at $\ell\approx\SI{1.2}{\micro\meter}$. As the thickness of the dispersive coating is reduced the two equilibrium points become more distant. This effect is better depictured in \Fig{Fig_4}(d), where the Casimir pressure is shown for a range of distances for four different coating thicknesses. In some cases (for instance when $d_1$ is smaller than $\SI{0.1}{\micro\meter}$, we find that one of the equilibrium points is at a distance which is too close to be practically accessible. However, one way to exchange the stability of these zero-pressure points would be to remove the dispersive coating of the PMC and coat the PEC mirror with a permeability-dispersive layer instead.
 
%  \begin{figure}[htpb]
%  \centering
%  \includegraphics[trim=0cm 0cm 0cm 0cm, clip=true,width=\columnwidth]{Fig6}
%  \caption{pressure dependence on the plasma frequency of the coating and separation for a cavity formed by a highly conductive wall and a coated high-impedance ($d_1=\SI{0.15}{\micro\meter}$ and $d_2=0$). Positive (negative) sign is associated with a repulsive (attractive) pressure.}
%  \label{Fig_6}
%  \end{figure}

\section{conclusion}
In this paper we have theoretically addressed the problem of the Casimir pressure between impedance boundaries (high and low impedances) with the addition of dielectric layers whose permittivity is frequency dispersive in terms of a Drude model. We have shown that when the plasma frequency is near the wavelength associated with the distance between the plates, the Casimir pressure can be designed by properly tuning the plasma frequency. Building on that rationale we have shown that when two different boundaries are included (a cavity made of a combination of high and low impedance walls), the pressure can be tuned between repulsive and attractive, with a zero-pressure crossing in between. We have found that those equilibrium points can be designed to be either stable or unstable at either close or far distances between the plates when compared to the thickness of the coating layer and their locations be tuned by varying the plasma frequency of the coating

\section{Acknowledgement}
The authors wish to acknowledge financial support from the Defense Advanced Research Program Agency (DARPA) QUEST program grant No. HR00112090084. The work of MC was partially funded by a postdoctoral fellowship from Junta de Andalucía (PAIDI) co-funded by the European Social Fund and by the Grant number PID2020-116739GB-I00 funded by MCIN/AEI/10.13039/501100011033.
%\nocite{*}
%

\end{document}